\documentclass[prl,twocolumn,aps,showpacs]{revtex4}
\usepackage{hyperref}
\usepackage{bm}
\def\ltsim{\vbox {\hbox{\lower .8\baselineskip \hbox{$<$}} \break
                 \hbox{\lower 0.2\baselineskip \hbox{$\sim$}} } }

\begin{document}

\title{Unification of electromagnetic noise and Luttinger liquid via 
a quantum dot}

\author{Karyn Le Hur and Mei-Rong Li}
\affiliation{D\'epartement de Physique, Universit\'e de Sherbrooke, Sherbrooke, 
Qu\'ebec, Canada J1K 2R1}

\begin{abstract}
We investigate the effect of dissipation on a small quantum dot (resonant 
level) tunnel-coupled to a chiral Luttinger liquid (LL) with the LL 
parameter $K$. The dissipation stems from the coupling of the dot to an 
electric environment, being characterized by the resistance $R$, via Coulomb 
interactions. We show that this problem can be mapped onto a 
Caldeira-Leggett model where the (ohmic) bath of harmonic oscillators 
is governed by the effective dissipation strength $\alpha=(2\tilde{K})^{-1}$ 
with $\tilde{K}^{-1}=K^{-1}+2R/R_K$ and $R_K=h/e^2$ the quantum of resistance. 
Experimental consequences are discussed and the limit $K=1/2^+$ is thoroughly 
studied at small $R/R_K$ through the spin-boson-fermion model.
\end{abstract}

\date{\today}

\maketitle

\begin{figure}[h]
\begin{picture}(250,120)
\leavevmode\centering\includegraphics{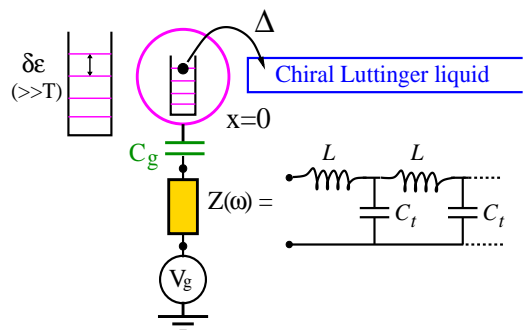}
\end{picture}
\caption{(Color online) A small dot (with large energy level 
spacing $\delta\epsilon$) coupled to a CLL ({\em e.g.}, a quantum wire 
at the edge $x=0$). The gate voltage exhibits 
fluctuations through an impedance $Z(\omega)$ modeled as an 
$LC_t$ transmission line.} 
\label{setup}
\end{figure}

A quantum dot can be viewed as a simple artificial atom exhibiting charge 
quantization \cite{Ashoori}, and its charge can now be measured with a very 
high accuracy with the aid of an electrometer based on a single-electron 
transistor \cite{Lehnert}. The coupling
of the quantum dot to a macroscopic reservoir of electrons inevitably 
produces quantum charge fluctuations on the dot.  This generic phenomenon has
been vividly investigated theoretically both in the case of a large 
metallic box with a very dense spectrum \cite{Matveev} and in the opposite 
limit of a two-level system \cite{Markus}. The reservoir of electrons may 
be a two-dimensional (2D) Fermi-liquid lead \cite{Matveev,Georg} or a 
1D structure \cite{Furusaki,Johan} [{\it e.g.},  a fractional quantum Hall 
edge state (FQHES)or a quantum wire] where interacting electrons form 
Tomonaga-Luttinger liquid (LL). Some recent endeavors have been accomplished 
by Cedraschi {\it et al.} \cite{Markus} and by one of us \cite{karyn} to 
understand the role of dissipation---coming from the capacitive coupling of the 
dot to an electric environment with an ohmic resistance---on the charge 
quantization of a quantum dot, but with the limitation of free electrons in 
the reservoir lead. In this Letter, we explore dissipation effects on the 
small quantum dot (resonant level) coupled to a chiral LL (CLL) which has been 
previously introduced by Furusaki and Matveev \cite{Furusaki}. We seek 
to provide a unified picture of the role
of interactions in the one-channel conductor  and of the zero-point 
fluctuations of the electric environment generalizing the case of a single 
junction \cite{Ines}. Of interest to us is to understand the nature of the 
quantum phases emerging through the dissipative mesoscopic structure shown in 
Fig.~\ref{setup} and hence to discuss physical implications for the occupation 
probability on the quantum dot. We highlight that the physics explored here is 
sufficiently appealing to experimentalists to carry out activities similar to 
those already existing on superconducting qubits capacitively coupled to lossy 
transmission lines in GaAs/AlGaAs heterostructures \cite{SCqubit}. 

The quantum dot of interest in Fig. 1 is small enough such that we can only 
restrict ourselves to the highest-occupied level. This leads to a two-level system 
\cite{Markus}. The gate voltage $V_g$ is fixed such that the two states in which 
the level is occupied ($|1\rangle$) or not ($|0\rangle$)
are almost degenerate. 
We can thus resort to an orbital spin-1/2 operator, ${\bf S}$, to describe these 
two states: $S_z=1/2$ corresponds to the state 
 $|1\rangle$ and $S_z=-1/2$ to 
$|0\rangle$;  $S^+$ flips the state from $|0\rangle$ to $|1\rangle$, and $S^-$ 
vice versa. In the presence of electromagnetic noise, the charging Hamiltonian 
takes the form
\begin{equation}
H_{c}=\epsilon S_z+({e\hat{Q}_0/C_g}) S_z. \label{charging}
\end{equation}
The detuning $\epsilon$ depending on the gate voltage $V_g$ is the energy 
difference that an electron must overcome if it wants to tunnel
between the dot and the lead; here we concentrate on the region close to 
$\epsilon=0$ (resonant level).
The second term in Eq.~(\ref{charging}) arises from the extra capacitive 
coupling between the dot and the gate voltage
fluctuations (the quantum noise) $\delta V_g(t)=\hat{Q}_0/C_g$, with $\hat{Q}_0$ 
denoting the charge fluctuation operator on the gate capacitor $C_g$ emerging from 
the finite impedance $Z(\omega)$ \cite{Markus,karyn}. This term has not been
previously considered in Ref.~\cite{Furusaki}. Akin to Refs.~[\onlinecite{Markus}] 
and [\onlinecite{LL}], we find appropriate to model the impedance $Z(\omega)$ in a 
microscopic fashion through a long dissipative transmission line composed of an 
infinite collection of $LC_t$ oscillators (Fig.~1); assuming $C_t=C_g$ 
\cite{noisenote}, the latter being governed by the Hamiltonian:
\begin{eqnarray}
H_{noise}= \int^1_0 dx \bigg\{ {\hat{Q}^2(x)\over 2C_t} + 
{\hbar^2 \over e^2} {2\over L} \sin^2\bigg({\pi x \over 2}\bigg) \hat{\phi}^2(x) 
\bigg \}, \label{Hnoise}
\end{eqnarray}
where the charge (fluctuation) operator $\hat{Q}(x)$ and the phase operator 
$\hat{\phi}(x)$ obey the commutation relation
$[\hat{\phi}(x),  \hat{Q}(y)/e] 
= i \delta(x-y)$. According to Ref.~[\onlinecite{LL}], $\hat{Q}_0 = \sqrt{2} 
\int^1_0 dx \cos (\pi x/2) \hat{Q}(x)$. At low frequency $\omega\ll 
\omega_c=1/(RC_t)$ where the resistance $R=\sqrt{L/C_t}$, the transmission line 
gives an impedance $Z(\omega)=R/(1+i\omega/\omega_c)\approx R$.

Now, we allow an electron to tunnel between the resonant level and the reservoir 
lead, {\it i.e.}, the CLL. Even though the CLL model is the most natural 
description of the FQHES \cite{Wen} the charge of the edge excitations depend 
sensitively on how contacting is done \cite{Chamon}. Therefore, semi-infinite 
quantum wires --- where the wires are coupled to the level only at the edge $x=0$ 
\cite{Furusaki,Tsvelik} as shown in Fig.~\ref{setup} --- represent a more judicious 
realization of the CLL for our proposal. We only consider the case of spinless 
electrons which implies that all the electrons have been completely spin-polarized 
by applying an external magnetic field. The kinetic part of the CLL reads 
\begin{equation}
H_{chiral}=\frac{v}{4\pi}\int_{-\infty}^{+\infty} \left(\frac{d\varphi}{dx}
\right)^2 dx,       \label{Hchiral}
\end{equation}
where $v$ is the Fermi velocity and the chiral boson field $\varphi(x)$ obey the 
commutation relations $[\varphi(x),\varphi(y)]=i\pi\hbox{sgn}(x-y)$. The tunneling 
processes between the CLL and the level can be described by the Hamiltonian 
\cite{Furusaki}
\begin{eqnarray}
H_{\Delta} = (\Delta/\sqrt{2\pi a}) 
\big(e^{i\varphi(0)/\sqrt{K}}S^+ + {\rm H.c.}\big), 
\label{Htunneling}
\end{eqnarray}
where $a$ is a short-distance cutoff, $\Delta$ the tunneling amplitude, 
$K<1$ the LL parameter, and we have exploited the bosonized form 
$\Psi(0)=(1/\sqrt{2\pi a}) \exp(i\varphi(0)/\sqrt{K})$ of an electron 
operator $\Psi(0)$ at $x=0$. In the case of the FQHES, $K$ must
be clearly identified as the Landau level filling factor \cite{Wen}. 
The total Hamiltonian 
takes the form $H_{tot}=H_c+H_{noise}+H_{chiral}+H_{\Delta}$. We make 
the unitary transformation ${\cal U}_1=\exp\{iS_z \hat{\phi}_0\}$, where 
$\hat{\phi}_0= \sqrt{2} \int^1_0 
dx \cos (\pi x/2)\hat{\phi}(x)$ is the conjugate operator to $\hat{Q}_0/e$, 
such that the noise contribution in $H_c$ is completely absorbed in the 
tunneling part as
\begin{equation}
\hskip -0.3cm \bar{H}_{\Delta}={\cal U}_1^{\dagger}H_{\Delta}{\cal U}_1=
(\Delta/\sqrt{2\pi a}) \big(e^{i\hat{\phi}_0}e^{i\varphi(0)/\sqrt{K}}S^+ 
+{\rm H.c.} \big).
\end{equation}

Apparently, the tunneling of an electron between the dot and the CLL must be 
mediated by excitations in the environmental bosonic modes.
It is crucial to bear in mind the large time behavior \cite{noise} 
\begin{equation}
{\cal K}(t) = \ \langle\hat{\phi}_0(t)\hat{\phi}_{0}(0)
\rangle - \langle {\hat{\phi}_0}^2 \rangle\ 
\simeq -2r\ln(i\omega_c |t|),
\label{Kt}
\end{equation}
where $r=R/R_K$ with $R_K=h/e^2\simeq 25.8k\Omega$ being the quantum of resistance. 
Let us first establish the renormalization group (RG) equation for the 
{\it dimensionless} tunneling amplitude $\tilde{\Delta}=\Delta/\sqrt{\Lambda}$; 
$\Lambda=\hbox{min}(\omega_c,\delta \epsilon)$ is the high-energy cutoff in our 
model, $\delta \epsilon$ the level spacing on the dot, and we must equate the 
frequency cutoff of the CLL to $v/a=\Lambda$ (we set $\hbar=k_B=1$ and $v$ is a 
dimensionless parameter). Expanding the partition function to second order in 
$\Delta$ and using ${\cal K}(t)$ in Eq.~(\ref{Kt}) give \cite{note}
\begin{equation}
d\tilde{\Delta}/d\hbox{l}=\big[1-(2K)^{-1}-r\big]\tilde{\Delta}, \label{RG}
\end{equation}
where the RG variable is $\hbox{l}=\ln(\Lambda/T)$ with $T\ll \Lambda$ denoting 
the temperature. This already allows us to distinguish two different regimes 
according to the parameter $\tilde{K}$ defined as $1/\tilde{K}=1/K+2r$.  For 
$\tilde{K}\ll 1/2$, $\tilde{\Delta}$ is an irrelevant perturbation which means that 
the physics is dominated by a level being weakly coupled to the CLL, 
and a perturbation theory in $\Delta$ is appropriate. This stands for the 
``localized'' phase where the level is occupied for $\epsilon<0$ and unoccupied 
for $\epsilon>0$, resulting in a jump in the occupation probability 
$\langle S_z\rangle_{\epsilon}$ of the level ($\epsilon$ is the electron 
energy relative to the Fermi energy) at 
$\epsilon=0$. When $\tilde{K}\gg 1/2$, the level coupling 
to the CLL is a relevant perturbation that will lift the degeneracy 
of the ground state at $\epsilon=0$ and lead to a continuous function of
$\epsilon$ for $\langle S_z\rangle_{\epsilon}$ 
\cite{Markus,Furusaki}. This is the ``delocalized" realm 
where an electron is resonating back and forth between the CLL and the dot. 
To get a better description of the strong-coupling fixed point for 
$\tilde{K}\gg 1/2$ as well as to investigate $\langle S_z\rangle_{\epsilon}$ 
on a more quantitative level, we derive an effective Caldeira-Leggett (or ohmic 
spin-boson) theory \cite{CL}. 

Noting that the level orbital spin only couples to the local CLL mode 
$\varphi(0)$ and to the ``local'' noise mode $\hat{\phi}_0$, it is convenient
to build the local actions for the modes $\varphi(\tau)=\varphi(x=0,\tau)$ and 
$\hat{\phi}_0(\tau)$ along the lines of Ref.~\cite{Tsvelik}:
\begin{eqnarray}
S^{loc}_{chiral} &=& (T/2\pi) \sum_{\omega_n} 
|\omega_n| \varphi(\omega_n)
\varphi(-\omega_n),   \label{action}\\ 
\nonumber
S^{loc}_{noise} &=&  (T/2\pi) \sum_{\omega_n}  |\omega_n| (2r)^{-1} 
\hat{\phi}_0(\omega_n)\hat{\phi}_0(-\omega_n),
\end{eqnarray}
where $\omega_n$ is the bosonic Matsubara frequency. 
Redefining the fields $\varphi_s = \sqrt{\tilde{K}}(\sqrt{K^{-1}} 
\varphi + \sqrt{2r} \, \bar{\phi}_0)$ and $\varphi_a = \sqrt{\tilde{K}}
(\sqrt{2r} \, \varphi - \sqrt{K^{-1}} \bar{\phi}_0)$ where 
$\bar{\phi}_0=(1/\sqrt{2r})\hat{\phi}_0$, we find that 
now the level gets only coupled to the mode $\varphi_s$, and therefore
it is sufficient to write down the action for $\varphi_s$ 
\begin{eqnarray}
S_{\varphi_s} &=& S_{\varphi_s}^{loc}+\bar{S}_{\Delta}= {T\over 2\pi} 
\sum_{\omega_n} |\omega_n| 
\varphi_s(\omega_n) \varphi_s(-\omega_n) \nonumber \\ 
&& \;\;\;\;-{\Delta}\sqrt{\Lambda/2\pi v}\, \big(e^{ i{\varphi_s}(0)
/{\sqrt{\tilde{K}}}}S^+ +H.c.\big).
\end{eqnarray}
Conceptually, we can visualize $S_{\varphi_s}$ as the action linked to an 
Hamiltonian $H_{\varphi_s}$ modeling a fictive CLL, which is described by an 
Hamiltonian $H_{chiral}\{\varphi_s\}$ similar to Eq.~(\ref{Hchiral}) with a LL 
parameter ${\tilde{K}}$, being coupled to the level. 
The link with the Caldeira-Leggett model of a two-level system with 
ohmic dissipation \cite{CL} appears when applying the unitary transformation 
${\cal U}_2=\exp\{i\varphi_s(0)S_z/\sqrt{\tilde{K}}\}$:
\begin{figure}[h]
\begin{picture}(250,105)
\leavevmode\centering\includegraphics{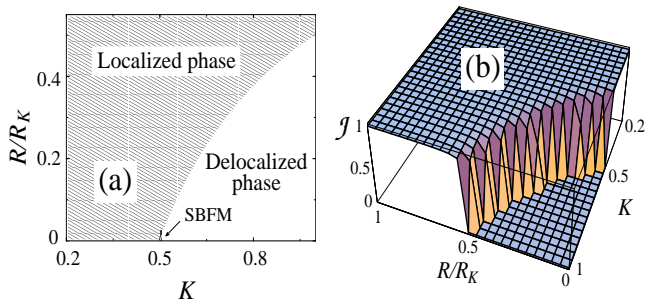}
\end{picture}
\caption{(Color online) (a) Our phase diagram.
The demarcation line is the $R_c/R_K$ versus $K$ curve. This generalizes 
the (noiseless) situation $R=0$ of Ref.~\cite{Furusaki}. 
(b) ${\cal J}=\langle S_z\rangle_{\epsilon=0-}-
\langle S_z\rangle_{\epsilon=0+}$ 
as a function of $K$ and $R/R_K$. Note the emergence of a conspicuous 
jump in the entrance of the localized phase. We have set
$\frac{\tilde{\Delta}^2}{2\pi v\Lambda}=0.01\ll 1$. Below the boundary,
${\cal J}=(2\tilde{K})^{1/2}\simeq 1$. 
} 
\label{phasediagram}
\end{figure}
\vspace{-0.2cm}
\begin{equation}
 {\cal U}_2^{\dagger}H_{\varphi_s}{\cal U}_2=H_{chiral}\{\varphi_s\}
+{\Delta}\sqrt{\frac{\Lambda}{2\pi v}}S_x-\frac{v}{2\sqrt{\tilde{K}}}S_z
\frac{d\varphi_s(0)}{dx}.
\end{equation}
It is well established in this model that  a Kosterlitz-Thouless (KT) type quantum 
phase transition, between a phase where a particle is confined in one of the two 
levels and another phase where it is delocalized, takes place at the dissipation 
strength $\alpha=1$ \cite{CL,Bray,Bulla}. Here, $\alpha=1/(2\tilde{K})$ remarkably 
consistent with our previous RG result. Note that $\tilde{K}\leq K$. 
The successful mapping thus immediately allows us to extract  a critical value of 
the resistance in the external circuit $R_c$ which is obviously a function of the 
strength of the interactions in the CLL:
 \begin{equation}
 R_c=[1-(2K)^{-1}]R_K
 \end{equation}
for $K\geq 1/2$, and $R_c=0$ otherwise. 
The resulting phase diagram is shown in Fig.~\ref{phasediagram}~(a). 
The border between the two phases marks the $R_c/R_K$ vs $K$ curve.
A significant physical consequence is displayed
in Fig.~\ref{phasediagram}~(b) as we discuss below. Fig.~\ref{phasediagram}
is the major result of this Letter. 

Now we study the quantity
${\cal J}=\langle S_z\rangle_{\epsilon=0^-}-\langle S_z\rangle_{\epsilon=0^+}$.
In the localized phase $(\tilde{K}< 1/2)$, we use a 
second-order perturbation theory in $\tilde{\Delta}\ll 1$ following 
Ref.~\cite{Markus} to get
\begin{equation}
\hskip -0.1cm  {\cal J}=1-\frac{\tilde{\Delta}^2}{2\pi v\Lambda}
\frac{{\tilde{K}}^2}{(1-\tilde{K})(1-2\tilde{K})}.
 \end{equation}
Here, $\tilde{\Delta}$ is a function of $T$ following the RG flow equation (\ref{RG}).
Very close to the phase transition region, $\tilde{\Delta}\simeq \Delta$ remains
approximately unrenormalized. 
We plot ${\cal J}$ as a function of $K$ and $r$ in 
Fig.~\ref{phasediagram}~(b) (the upper stair region). Here, the occupation 
probability $\langle S_z\rangle_{\epsilon}$ displays a flagrant discontinuity at 
the Fermi level reflecting the charge quantization on the quantum dot: for 
$\epsilon<0$ the level is fully occupied whereas for $\epsilon>0$ this is empty.
From Eq.~(9), it becomes transparent that the suppressed noise spectrum in 
the electric environment at low energy contributes to reinforcing
the vanishing of the tunneling density of states in the CLL at 
$\epsilon=0$ \cite{note2,Furusaki}, $\rho(\epsilon)\propto \epsilon^{\mu}$ with 
$\mu=1/\tilde{K}-1=1/K+2r-1 > 1$, which 
results in a visible localization of particles in the CLL at the edge $x=0$ 
(and on the dot) over a large domain of the phase diagram. Just below the 
boundary, we can generalize the arguments of Ref.~\cite{Furusaki} to reach a 
critical value of the jump ${\cal J}=(2\tilde{K})^{1/2}$ which is almost 
{\it one}. 

In the delocalized phase, the low energy properties of the Caldeira-Leggett model 
can be identified to those of the anisotropic Kondo model with the Kondo couplings 
$J_{\perp}=\Delta\sqrt{\Lambda/2\pi v}$ and 
$J_z={v\over 2}[1-1/(2\tilde{K})^{1/2}]$ \cite{Furusaki}. Note that the analogy 
between the CL problem and the Kondo model was first pointed out in Refs. 
\onlinecite{Bray}. Not too far from the border 
({\it i.e.}, ${\tilde{K}}-1/2\ll 1$), we are in the weak-coupling Kondo limit
where one can easily make use of Bethe-Ansatz results \cite{Tsvelik2}. 
The emerging Kondo scale $\Gamma$ corresponds explicitly to the energy scale 
at which the coupling $\tilde{\Delta}$ gets strongly renormalized in Eq.~(7):
$\Gamma=\Lambda\tilde{\Delta}^{2\tilde{K}/(2\tilde{K}-1)}.$
The Kondo ground state for energies smaller than $\Gamma$ can be viewed as 
the complete screening of the orbital spin $S_z$ in the absence of magnetic 
field $(\epsilon=0)$ implying $\langle S_z \rangle_{\epsilon=0^\pm}=0$; 
equivalently, the prominent tunneling process smears the quantization of 
charge on the dot, {\it i.e.}, makes the occupation probability
$\langle S_z\rangle_{\epsilon}$ continuous around the Fermi level, and thus 
${\cal J}=0$ [shown as the lower stair region in Fig.~\ref{phasediagram}~(b)]. 
This is a clear distinction between the localized and delocalized phases which 
in principle should be accessible experimentally \cite{Lehnert}. 

The case of $K=1/2^+$ deserves a special treatment. We still 
resort to the original Hamiltonian $H_{tot}$. For $r\ll 1$, the tunneling 
process in Eq.~(4) becomes a marginal operator such that  higher order terms 
in $\Delta$ will play a role in the RG flow of
Eq.~(7). This will slightly modify the value of $R_c$. We resort to the 
spin-boson-fermion model (SBFM) or equivalently the Bose-Fermi Kondo model
 \cite{BF} to make a zoom into this area denoted by SBFM in 
Fig.~\ref{phasediagram}~(a). We refermionize $H_{\Delta}$ in 
Eq.~(\ref{Htunneling}):
\begin{equation}
H_{\Delta}= J_{\perp}
\left(\psi^{\dagger}_{\downarrow}(0)\psi_{\uparrow}(0)S^+ +
{\rm H.c.}\right),
\end{equation}
with the dimensionless Kondo coupling $J_{\perp}=\Delta\sqrt{2\pi v/\Lambda}$ 
and $\psi^{\dagger}_{\downarrow}(x)\psi_{\uparrow}(x)=[\Lambda/(2\pi v)]
\exp(i\sqrt{2}\varphi(x))$. Now, let us rewrite $H_c$ in Eq.~(\ref{charging}) 
exactly like in Ref.~\cite{karyn}:
\begin{equation}
H_c=(\epsilon+\sqrt{r}\Phi) S_z + J_z S_z \left(\psi^{\dagger}_{\uparrow}(0)
\psi_{\uparrow}(0)-\psi^{\dagger}_{\downarrow}(0)\psi_{\downarrow}(0)\right),
\end{equation}
where $\Phi=e\delta V_g/\sqrt{r}$ represents the bosonic variable coupled to the 
level, and to be fully consistent we have included the Ising part $J_z$ of the 
Kondo coupling which emerges in the renormalization procedure or which might 
also be induced by a small Coulomb interaction $uS_z d\varphi(0)/dx$ between 
electrons in the CLL and the quantum dot. Thus far we have assumed $u=0$, so at the 
bare level $J_z=0$. According to Refs.~\cite{karyn,LL}, we can then predict that the 
delocalized-localized transition will occur at a {\it finite} but small $R_c$; in 
the limit of very small $J_{\perp}$, we rigorously obtain $R_c=R_K J_{\perp}$. We 
finally expect a small jump in the value of $R_c$ at $K=1/2$ as depicted in 
Fig.~2 (a). A quantitative discussion for larger values of ${\Delta}$ including a 
finite Coulomb repulsion $u$ will be addressed elsewhere through a numerical 
RG approach \cite{NRG}.

The above analysis can also be generalized to the situation where the small dot
in Fig.~\ref{setup} is replaced by a large metallic grain with 
$\delta\epsilon\rightarrow 0$. We can
assume that such a metallic dot is similar to a small 
quantum wire governed by a Fermi liquid theory, {\it i.e.}, with a LL 
parameter $K_g=1$; charging effects are again taken into account through Eq. (1)
but now $S_z$ must be viewed rigorously as a projecting operator acting on the 
two {\it charge} states $Q=0$ and $Q=1$.
This is indeed a valid description when focusing on local 
physics around $x=0$ \cite{Matveev,Georg}. The tunneling Hamiltonian 
in Eq.~(\ref{Htunneling}) is replaced by 
$H_{\Delta'} = \Delta'\frac{\Lambda}{2\pi v} \big(e^{i\varphi(0)/\sqrt{K}}
e^{-i\varphi_g(0)}S^+ +h.c.\big),$
implying that now $\Delta'$ is a dimensionless quantity. The operator
$S^+$ flips the charge state $Q=0$  
on the grain to $Q=1$ \cite{Matveev,Georg}. We can introduce the 
boson fields $\varphi_1=\sqrt{K_1}(\varphi/\sqrt{K}-\varphi_g)$ and 
$\varphi_2=\sqrt{K_1}(\varphi+\varphi_g/\sqrt{K})$ such that the 
orbital spin is coupled to $\varphi_1$ only and the
free Hamiltonian 
for $\varphi_1$ takes exactly the same form as in Eq.~(\ref{Hchiral}) 
with the rescaled LL parameter $K_1$ defined as $1/K_1=1+1/K$. We can then 
proceed in 
a similar manner as before and anticipate a clear shift of the phase 
boundary in Fig.~\ref{phasediagram} which is now determined by the equality  
$2r+1/K_1=2r+1/K+1=2$. It is crucial to note that for moderate 
repulsive interactions in the CLL, meaning $K<1$, the system will be in
the localized phase already at $r=0$, {\it i.e.},
$R_c=0$, therefore we infer that the 
quantum noise has a 
minor effect on the results. Nevertheless, for a Luttinger exponent $K=1^-$ 
which should correctly mimic the situation of a 2D electron 
reservoir \cite{Georg}, we reproduce the SBFM 
introduced by one of us in Ref.~\cite{karyn}, and especially we recover 
that $R_c=R_K \Delta'$ for very small tunneling amplitudes 
$\Delta'$. 

To summarize succinctly, we have investigated dissipation effects on a 
small quantum dot coupled to a CLL which, {\em e.g.}, can describe a 
semi-infinite quantum wire as shown in Fig.~1. When the dissipation stems from 
a lossy transmission line we have been capable of providing a unified description 
of the role of the interactions in the CLL and of the zero-point fluctuations 
in the electric environment through the Kondo physics; the derived Caldeira-Leggett 
theory extending the previous works of Refs.~\cite{Markus,Furusaki,karyn}
allows us to predict a KT phase transition  at $\tilde{K}^{-1}=2R/R_K+K^{-1}=2$, 
separating a localized and a delocalized phase of the dot.  Increasing the size of 
the dot from the nanoscale to the micronscale would strongly reduce the delocalized 
realm. Recent advanced material technology allows to fabricate clean quantum wires 
with $0.5<K<1$ in GaAs/AlGaAs heterostructures \cite{Yacoby}. Other groups have 
used similar heterostructures to generate a superconducting quantum dot capacitively 
coupled to a 2D electron gas which serves as the dissipation bath \cite{SCqubit}. 
In particular, the characteristic $R$ can be tuned through changing the 
2D electron density. We thus anticipate that the setup shown in Fig.~1 can be 
built up in such semiconducting heterostructures. The noise induced quantum
phase transition we predicted here can thus be tested in experiments, especially
by noting that for $K\ll 1$, $R_c\ll R_K=25.8k\Omega$ which can be easily 
accessed by experiments.

K.L.H. is grateful to S.~Florens, K. Lehnert, and P.~Simon for valuable discussions 
on electric noise. This work was supported by CIAR, FQRNT, and NSERC.


\vspace{-0.3cm}

\begin{thebibliography}{10}

\bibitem{Ashoori} R. C. Ashoori, Nature {\bf 379}, 413 (1996).
\bibitem{Lehnert} D. Berman, N. B. Zhitenev, R. C. Ashoori and M. Shayegan, 
Phys. Rev. Lett. {\bf 82}, 161 (1999); 
K. W. Lehnert, B. A. Turek, K. Bladh, L. F. Spietz, D. Gunnarsson, P. Delsing, and 
R. J. Schoelkopf, Phys. Rev. Lett. {\bf 91}, 106801 (2003).
\bibitem{Matveev} K. A. Matveev, Zh. Eksp. Teor. Fiz. {\bf 99}, 1598 (1991) 
[Sov. Phys. JETP {\bf 72}, 892 (1991)].
\bibitem{Markus} P. Cedraschi, V. V. Ponomarenko, and M. B\"{u}ttiker, 
Phys. Rev. Lett. {\bf 84}, 346 (2000); P. Cedraschi and M. B\"{u}ttiker, Annals
of Physics {\bf 289}, 
1-23 (2001).
\bibitem{Georg} K. Le Hur and G. Seelig, Phys. Rev. B {\bf 65}, 165338 (2002).
\bibitem{Furusaki} A.~Furusaki and K.~A.~Matveev, Phys. Rev. Lett. 88, 226404 (2002). 

\bibitem{Johan} P. Kakashvili and H. J. Johannesson, Phys. Rev. Lett. {\bf 91}, 
186403 (2003); E. H. Kim, Y. B. Kim, and C. Kallin, J. Phys. condens. matt. 
{\bf 15}, 7047 (2003).
\bibitem{karyn} K. Le Hur, Phys. Rev. Lett. {\bf 92}, 196804 (2004).
\bibitem{Ines} I. Safi and H. Saleur, Phys. Rev. Lett. {\bf 93}, 126602 (2004);
M. Sassetti and U. Weiss, Europhys. Lett. 27, 311 (1994).
\bibitem{SCqubit} J. B. Kycia {\it et al.}, Phys. Rev. Lett. {\bf 87}, 017002 
(2001); A. J. Rimberg and W. Lu, cond-mat/0205382.
\bibitem{LL} Mei-Rong Li and Karyn Le Hur, Phys. Rev. Lett. {\bf 93}, 176802 (2004).
\bibitem{noisenote} This assumption is required for the exact diagonalization of the 
model for the transmission lines. 
For a generic case of $C_t\neq C_g$, there is an extra 
term in the noise Hamiltonian $(1/2C_g-1/2C_t)\hat{Q}_0^2$, and one would thus 
not reach the Johnson-Nyquist nois local action $S^{loc}_{noise}$ shown in 
Eq.~(\ref{action}).
On the other hand, the resistance $R$ is the only relevant physical parameter
and the transmission lines are purely a phenomenological model for it.
Therefore, $R$ leads to the Johnson-Nyquist noise which can be produced from the 
transmission lines only for $C_t=C_g$. 
\bibitem{Wen} X.-G. Wen, Int. J. Mod. Phys. B {\bf 6}, 1711 (1992).
\bibitem{Chamon} C. de C. Chamon and E. Fradkin,  Phys. Rev. B {\bf 56}, 2012 (1997).
\bibitem{Tsvelik} A. O. Gogolin, A. A. Nersesyan, and A. M. Tsvelik, in 
{\it Bosonization and Strongly Correlated Systems}, Cambridge University Press, 
Cambridge 1999 (chapter 27).
\bibitem{noise} Yu. V. Nazarov, Sov. Phys. JETP {\bf 68}, 561 (1989);
Yu.~V. Nazarov and  G.-L. Ingold in {\it Single Charge Tunneling}, 
edited by H. Grabert and M.~H. Devoret (Plenum Press, New York, 1992), Chap. 2, 
pp. 21-107; M. H. Devoret {\it et al.}, Phys. Rev. Lett. {\bf 64}, 1824 (1990).
\bibitem{note} The correction to the partition function to second order in 
$\tilde{\Delta}$ reads $\delta Z \approx
-\tilde{\Delta}^2\Lambda/[(2\pi)^2 a]\int d\tau_1\int d\tau_2 
{\cal K}(\tau_1-\tau_2)(a/(v|\tau_1-\tau_2|))^{1/K}$ where we must equate 
$\omega_c=\Lambda=v/a$ and $\tau_i=it_i\gg 1/\Lambda$. 
$\delta Z$ must be independent from the energy cutoff $\Lambda$, which results in 
Eq. (7). 
\bibitem{CL} A. O. Caldeira and A. J. Leggett, Annals of Physics {\bf 149}, 374 
(1983); A. J. Leggett, S. Chakravarty, A. T. Dorsey, Matthew P. A. Fisher,
Anupam Garg, and W. Zwerger, Rev. Mod. Phys. {\bf 59}, 1 (1987).
\bibitem{Bray} S. Chakravarty, Phys. Rev. Lett. {\bf 49}, 681 (1982); 
A. J. Bray and M. A. Moore, Phys. Rev. Lett. {\bf 49}, 1545 (1982).
\bibitem{Bulla} R. Bulla, N.-H. Tong, and M. Vojta, Phys. Rev. Lett. {\bf 91}, 
170601 (2003). 
\bibitem{note2} The tunneling density of states of a LL at $x=0$ (CLL) takes the 
well-known form $\rho(\epsilon)\propto \epsilon^{(1/g) -1}$, $g$ being the LL 
exponent. From Eq. (9), we deduce $g=\tilde{K}$.
\bibitem{Tsvelik2} A. M. Tsvelik and P. B. Wiegmann, Adv. Phys. {\bf 32}, 453 
(1983); N. Andrei, K. Furuya, and J. H. Lowenstein, Rev. Mod. Phys. {\bf 55}, 331 
(1983).
\bibitem{BF} See, {\it e.g.}, L. Zhu and Q. Si, Phys. Rev. B {\bf 66}, 024426 
(2002); G. Zarand and E. Demler, Phys. Rev. B {\bf 66}, 024427 (2002).
\bibitem{NRG} M.-R. Li, K. Le Hur, and W. Hofstetter, Phys. Rev. Lett. {\bf 95},
086406 (2005).
\bibitem{Yacoby} O. Auslaender, A. Yacoby, R. de Picciotto, K. W. Baldwin,
L. N. Pfeiffer, and K. W. West, Science {\bf 295}, 825 (2002).

\end{thebibliography}
\end{document}